\documentclass[%
 reprint,
 amsmath,amssymb,
 aps,
 pra,
floatfix,
]{revtex4-2}

\usepackage{graphicx}
\usepackage{dcolumn}
\usepackage{bm}

\newcommand{\bra}[1]{\langle #1 |}
\newcommand{\ket}[1]{| #1 \rangle}

\begin{document}
\title{Macroscopic properties of high-harmonic generation from molecular ions}
\author{T. Joyce}
\affiliation{JILA and Department of Physics, University of Colorado, Boulder, USA}
\author{A. Jaron}
\affiliation{
J. R. Macdonald Laboratory, Physics Department, Kansas State University,
Manhattan, USA}
\affiliation{JILA and Department of Physics, University of Colorado, Boulder, USA}


\begin{abstract}

  We  extend  the  existing  framework  of macroscopic HHG to combine it with high-accuracy ab initio calculations for molecules as microscopic input. This approach  is  applied  to  HHG spectra exhibiting Mollow sidebands, for open shell molecules undergoing nonadiabatic dynamics. We demonstrate the details of the method and analyze how the predicted features in the microscopic HHG response unambiguously  survive  macroscopic  response  calculations,  and furthermore they exhibit a interesting angular pattern in the far-field.

We calculate the
macroscopic harmonic spectrum by combining many single-molecule calculations at different intensities, obtained in one case from time-dependent density functional theory calculations for N$_2^+$, in second case for one electron time dependent Schr\"odinger equation for a 1D double well model potential. For both cases one can observe that the resulting macroscopic spectra exhibit  Mollow sidebands of approximately the same intensity as the main harmonics, while being radiated at wider angles,
meaning they could be isolated more easily in an experiment.
\end{abstract}

\maketitle

\section{Introduction}
High harmonic generation (HHG) is a highly nonlinear process where many low energy photons are converted into a single high energy photon.
Because the process happens within only a few attoseconds, the measured harmonic spectrum contains an imprint of attosecond dynamics induced by an intense laser pulse.

In molecular ions, it was recently predicted that the usual odd harmonics can be accompanied by sidebands at non-integer multiples of the fundamental frequency, intense field  analog of Mollow triplets in quantum optics\cite{Xia16,TJ20}. 
The Mollow triplet, first described theoretically by B. R.Mollow in 1969  \cite{mollow}, is structure of three peaks  in the fluorescence spectrum for a two-level system. It has been 
observed in many different systems such as for example: ions \cite{stalgies}, single molecules \cite{wrigge}, or quantum dots \cite{uhlaq,flagg,peiris,lagoudakis}.
Importantly it is considered a fundamental signature of quantum optics, due to photon correlations between the sidepeaks of the spectrum, first reported in  \cite{aspect}. Later this topic has attracted more attention for example as a promising candidate of heralded single-photon sources \cite{lopez}.

In present work we describe the Mollow sidebands in HHG that appear as a  result from competition between
two nonperturbative processes, Rabi oscillation and high-harmonic generation.
In this paper, we describe in detail and analyze a simple model for the effects of intensity averaging on macroscopic HHG spectra, and apply it to the case of Mollow sidebands in high harmonics from the nitrogen molecular ion. 
This paper is structured as follows.
In Sec. \ref{sec:macroeq} we introduce a model for  macroscopic harmonic generation.
In Sec. \ref{sec:methods} we describe our numerical method for both the ab initio calculations of single molecule harmonic spectra, and our implementation of the macroscopic intensity averaging procedure.
In Sec. \ref{sec:results} we discuss the results of our calculations of macroscopic harmonic spectra.


\section{Calculations of the macroscopic HHG response}
\label{sec:macroeq}

The driving laser  induces a polarization density $\mathbf{P}(\mathbf{r},\omega)$ in the medium, from which we would like to calculate the far-field radiation spectrum $U(\omega,\hat{n})$ as a function of angular frequency $\omega$ and direction $\hat{n}$.

A standard electrodynamics calculation gives the following result
\begin{equation}
\label{eq:fraunhofer}
U(\omega,\hat{n}) = \frac{\omega^4}{2 c^3}\left|\int \hat{n}\times\mathbf{P}(\mathbf{r},\omega) e^{i(\omega/c)\hat{n}\cdot\mathbf{r}}d^3\mathbf{r}\right|^2.
\end{equation}
Next, we assume that the laser is a Gaussian beam of width $b$ propagating in the $\hat{z}$-direction, and polarized in the $\hat{x}$ direction.
Moreover we assume that the size of the medium is much less than the Rayleigh length of the beam, and that its focus lies in the medium.
Consequently the laser's electric field at a macroscopic location $\mathbf{r}=(x,y,z)$ and time $t$ is
\begin{align}
\label{eq:gaussianBeam}
\boldsymbol{\mathcal{E}}(\mathbf{r},t)=
\hat{x}\mathcal{E}_0 e^{-(x^2+y^2)/b^2}
f\left(t - \frac{z}{c}\right),
\end{align}
for some envelope function $f$.
Within the dipole approximation, the induced polarization at each point is
\begin{equation}
\label{eq:polar}
\mathbf{P}(\mathbf{r},\omega) = \hat{x}\rho(\mathbf{r}) e^{-i\omega z/c} D(\mathcal{E}_0e^{-(x^2+y^2)/b^2},\omega),
\end{equation}
where $\rho(\mathbf{r})$ is the macroscopic number density of molecules near position $\mathbf{r}$ and $D(\mathcal{E},\omega)$ is the polarization induced in a single molecule by an electric field with peak amplitude $\mathcal{E}$ and envelope $f$.
Substituting Eq. (\ref{eq:polar}) into Eq. (\ref{eq:fraunhofer}) gives the macroscopic equation for an arbitrary target geometry,
\begin{eqnarray}
\label{eq:macrogeneral}
U(\omega,\hat{n}) &=& \frac{\omega^4}{2 c^3}|\hat{n}\times\hat{x}|^2
\Big|\int \rho(\mathbf{r})e^{i(\omega/c)(\hat{n}-\hat{z})\cdot\mathbf{r}} \nonumber \\
&\times& D(\mathcal{E}_0e^{-(x^2+y^2)/b^2},\omega)d^3\mathbf{r} \Big|^2.
\end{eqnarray}
In particular, the spectrum in the forward direction is
\begin{equation}
\label{eq:forward}
U(\omega,\hat{z})  = \frac{\omega^4b^4}{8 c^3}\left|\int_0^{\mathcal{E}_0} D(\mathcal{E},\omega)N\left(\mathcal{E}\right)\frac{d\mathcal{E}}{\mathcal{E}}\right|^2,
\end{equation}
i.e., a coherent sum of the microscopic spectra weighted by a geometry dependent factor
\begin{align}
\label{eq:forwardN}
    N(\mathcal{E})=\frac{\int_{-\infty}^\infty \rho(\mathbf{r})\delta(\mathcal{E}-\mathcal{E}_0e^{-(x^2+y^2)/b^2})d^3\mathbf{r}}
    {\int_{-\infty}^\infty \delta(\mathcal{E}-\mathcal{E}_0e^{-(x^2+y^2)/b^2})d^3\mathbf{r}}.
\end{align}
Typically the oscillating factor in Eq. (\ref{eq:macrogeneral}) causes the integral to vanish unless $\hat{n}\approx\hat{z}$ (forward focusing).
Nonetheless, the angle resolved spectrum $U(\omega,\hat{n})$ can vary rapidly and nontrivially in the neighborhood of the forward direction, so it is not enough to only evaluate Eq. (\ref{eq:forward}).

In many experimental setups, the laser beam is narrow enough that the dependence of number density on the transverse coordinates $x$ and $y$ can be neglected: $\rho(\mathbf{r})=\rho(z)$.
We call this a slab geometry, because the target effectively has infinite extent in the $x$ and $y$ directions but a finite thickness in the $z$ direction characterized by $\rho(z)$.
This assumption about the target geometry allows us to simplify Eq. (\ref{eq:macrogeneral}) to a product of a geometric factor $G(\kappa)$ and a universal factor $K(\omega,\kappa)$ which is independent of both the geometry and the beam width, namely,
\begin{eqnarray}
\label{eq:u_slab}
U(\omega,\hat{n}) =& \frac{\omega^4b^4}{2 c^3}|\hat{n}\times\hat{x}|^2G\left(\frac{\omega}{c}\sin^2\frac{\theta}{2}\right)
K\left(\omega,\frac{\omega b}{c}\sin\theta\right) 
\end{eqnarray}
where $G$ and $K$ are defined as 
\begin{align}
\label{eq:gdef}
G(\kappa) =&  \left|\int_{-\infty}^\infty \rho(z)e^{i\kappa z}dz \right|^2,\\
\label{eq:hdef}
K(\omega,\kappa) =& \left|\int_0^\infty D(\mathcal{E}_0e^{-q^2},\omega) J_0\left(\kappa q\right)q dq\right|^2,
\end{align}
where $\theta=\cos^{-1}(\hat{n}\cdot\hat{z})$ and $J_0$ is the zeroth-order Bessel function.
The spectrum in the forward direction is
\begin{align}
    U(\omega,\hat{z}) = \frac{\omega^4b^4}{2 c^3} G(0)K(\omega,0),
\end{align}
which is equivalent to Eq. (\ref{eq:forward}) with $N(\mathcal{E})=\sqrt{G(0)}$,
and the angle-integrated spectrum is
\begin{eqnarray}
\label{eq:integrated}
    \int U(\omega,\hat{n})&d&^2\hat{n} = \frac{\omega^4b^4}{4 c^3}\int_0^\pi \ d\theta \sin \theta 
    \left[1+\cos^2\theta\right] \nonumber \\
    &\times& G\left(\frac{\omega}{c}\sin^2\frac{\theta}{2}\right)
K\left(\omega,\frac{\omega b}{c}\sin\theta\right). 
\end{eqnarray}


In our calculations we take the number density to have a Gaussian distribution,
\begin{align}
\label{eq:rhoz}
    \rho(z) = \rho_0e^{-z^2/(2z_0^2)}.
\end{align}
The thickness of the target $z_0$ should be much smaller than the Rayleigh length of the driving laser $z_0\ll z_R=\pi b^2/\lambda$, otherwise we would need to modify Eq. (\ref{eq:gaussianBeam}) to account for the $z$-dependence of $b$ and $f$.
Since we use a beam waist of $b=30~\mu$m and wavelengths of $\lambda=550$ nm and $\lambda=446$ nm (see Sec. \ref{sec:methods})---which give $z_R=5.1$ mm and $z_R=6.3$ mm respectively---we choose the thickness to be $z_0=0.5$ mm, which both satisfies the constraint by an order of magnitude and is an experimentally realistic thickness for a gas jet or other target.
We have tried different values for these macroscopic parameters and found that, at least for the cases studied in this paper, the results do not depend strongly on the values of macroscopic parameters---assuming they stay within the specified range such that the assumptions we have made in this section are valid.

\section{Numerical methods}
\label{sec:methods}
In this section, we detail our numerical methods for computing $H(\omega,\kappa)$ as defined in Eq. (\ref{eq:hdef}).
This is divided into two steps.
First, we explain how the single-molecule HHG response $D(\mathcal{E},\omega)$ is computed for a particular value of the peak electric field strength $\mathcal{E}$.
Second, we introduce a highly efficient numerical algorithm to compute the integral in Eq. (\ref{eq:hdef}) using only a very small number of direct evaluations of $D(\mathcal{E},\omega)$.

\subsection{Single-molecule calculations}
\label{sec:methods1}
The macroscopic equations derived in Sec. \ref{sec:macroeq} require as an input $D(\mathcal{E},\omega)$, which is the Fourier transform of the time-dependent dipole moment induced in the molecule.

Although in many cases approximate semi-classical models like Strong Field Approximation (SFA) and its variants give accurate results for HHG, it is known that SFA does not properly describe below-threshold harmonics, the effect of excited states on the harmonic spectrum and ellipticity of molecular harmonics (see e.g. \cite{YX-ellipt}). 
Since we are interested in computing the macroscopic response for system under conditions where we expect such features, we cannot use SFA and must instead solve the time-dependent Schrodinger equation (TDSE). 

First, we studied a 1D model system with a Gaussian double well representing a generic aligned diatomic molecule,
\begin{align}
    \hat{H}_{1D} = -\frac{1}{2}\frac{\partial^2 }{\partial x^2} - e^{(x-2)^2/2}-e^{(x+2)^2/2}.
\end{align}
The ground state $\Psi_g(x)$ is at $E_0=-0.642$ a.u. and the first excited state is at $E_1 = -0.559$ a.u. 
We choose the central frequency of the driving laser to be resonant with that transition $\omega_0=E_1 - E_0 = 0.083$ a.u., corresponding to a wavelength of 549 nm,
with a Gaussian envelope of full-width at half-maximum (FWHM) duration $\tau=400$ a.u., centered at $t_0 = 2\tau$,
\begin{align}
\label{eq:efield1D}
    f(t) = e^{-(t-t_0)^2/(2\tau^2)}\cos(\omega_0 t).
\end{align}
The peak electric field amplitude $\mathcal{E}$ is left as a free parameter which we will sweep over (see Sec. \ref{sec:methods2}).
For a particular value of $\mathcal{E}$, we get the time-dependent wavefunction $\Psi_\mathcal{E}(x,t)$ by solving the following TDSE,
\begin{equation}
    i\frac{\partial}{\partial t}\Psi_\mathcal{E}(x,t) =  \left[\hat{H}_{1D} + \mathcal{E}x f(t)\right]\Psi_\mathcal{E}(x,t),
\end{equation}
where the initial state for the time propagation is the ground state
\begin{equation}
    \Psi_\mathcal{E}(x,0)=\Psi_g(x) .
\end{equation}
Numerically, we discretize the Hamiltonian using fourth order finite difference with a grid spacing of $dx=0.3$ a.u. and a box size of 30 a.u. in each direction (total length 60 a.u.).
The last 5 a.u. in each direction are used for a complex absorbing potential.
We time propagate with the Crank-Nicholson method using a time step $dt=0.1$ a.u. from $t=0$ up to $t=2t_0$.
Once $\Psi_{\mathcal{E}}(x,t)$ is calculated, we compute the time dependent dipole moment using the formula,
\begin{align}
    D(\mathcal{E},t) = \int_{-\infty}^\infty |\Psi_{\mathcal{E}}(x,t)|^2xdx.
\end{align}
The Fourier transform in time is calculated
using the acceleration form 
\begin{align}
\label{eq:fttime}
    \omega^2 D(\mathcal{E},\omega) = \int_{0}^{2t_0} \frac{\partial^2}{\partial t^2} D_{1D}(\mathcal{E},t) W\left(\frac{t}{2t_0}\right)dt,
\end{align}
with a Blackman window $W$,
\begin{equation}
    W(\eta) = 0.42 - 0.5\cos(2\pi\eta) + 0.08\cos(4\pi\eta).
\end{equation}

\begin{figure}
    \centering
    \includegraphics[width=0.98\linewidth]{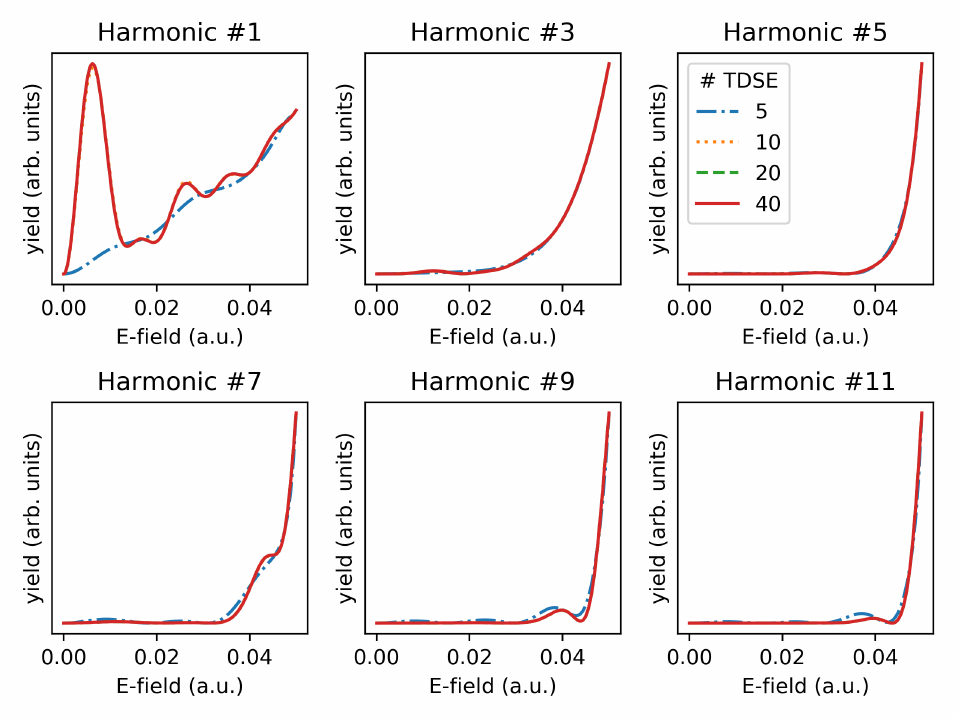}
    \caption{Convergence of the intensity-dependent single-molecule spectra, for selected harmonic frequencies, as a function of the number of TDDFT evaluations used for the interpolation (see legend in the top-right).
    Due to the rapid convergence of polynomial interpolation through Chebyshev nodes, the spectra interpolated using 10,20, and 40 TDSE evaluations are visually indistinguishable.
    }
    \label{fig:convergence}
\end{figure}

\begin{figure}
    \centering
\includegraphics[width=0.6\linewidth]{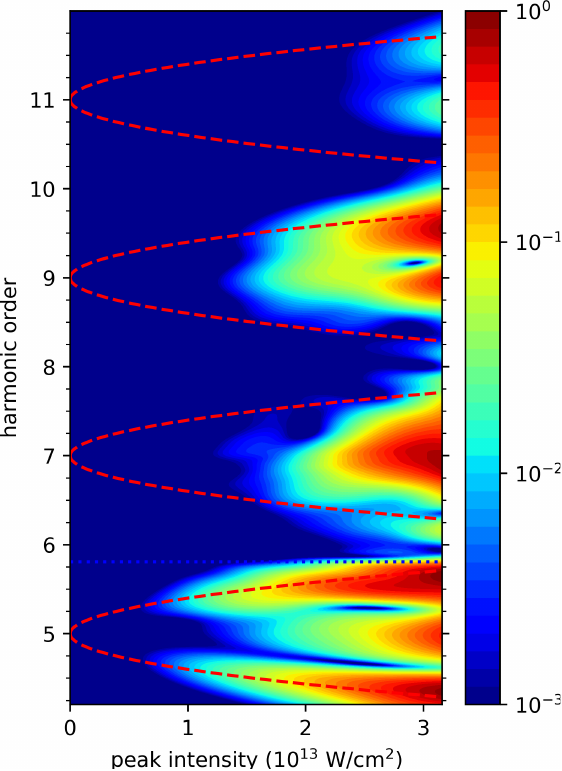}
   \caption{
    HHG spectra of a single molecule 1D model for different peak laser intensities.}  
    \label{fig:1Dmicroa}
\end{figure}

\begin{figure}
    \centering
\includegraphics[width=0.6\linewidth]{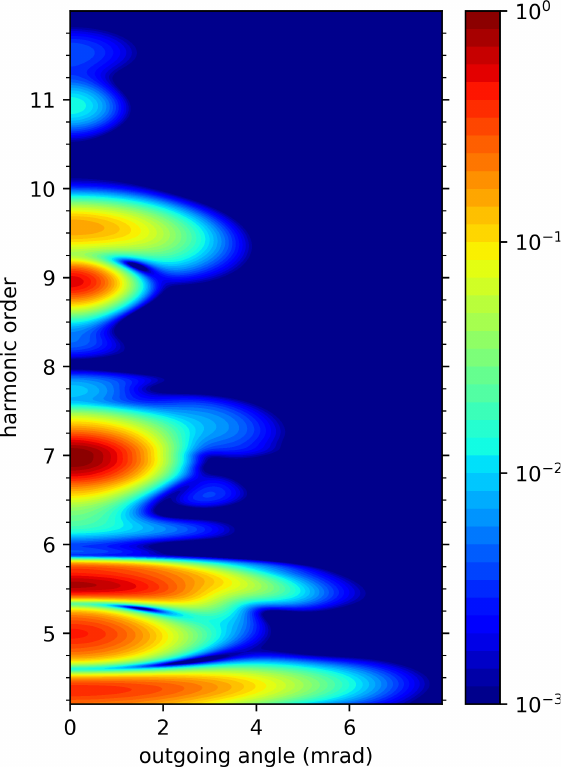}
    \caption{
      Macroscopic results for HHG spectra for 1D model molecule
    shown as function of the outgoing angle.
    Some of the features can be explained with an analytical (Floquet) dressed state model (dashed red lines). Blue dashed line marks excited state energy. }
    \label{fig:1Dmicrob}
\end{figure}

\begin{figure}[ht]
    \centering
    \includegraphics[width=0.6\linewidth]{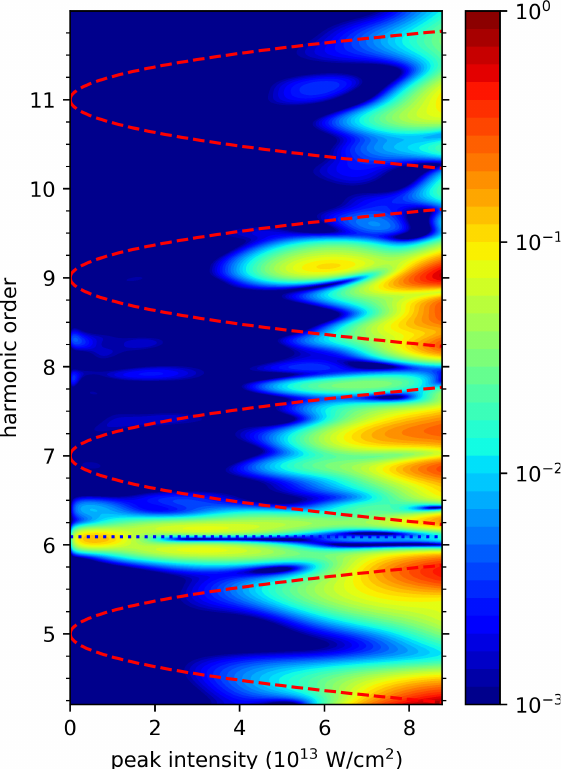}
    \caption{Theoretical HHG spectra of a single aligned $N_2^+$ molecule driven by a 446 nm laser as a function of peak electric field amplitude. Besides the usual odd harmonics, many other features appear, some of which can be explained using an analytical model based on Floquet theory (dashed red lines).
    The analytical model is very successful for low harmonics (1,3,5,6), and moderately successful for above-threshold harmonics (11,13), but fails just below the ionization threshold (7,9) where there is a more complicated structure due to Rydberg states.
    }
    \label{fig:N2micro}
\end{figure}

For N$_2^+$ in 3D, the calculations are performed within Time Dependent Density Functional Theory  (TDDFT) as implemented in the Octopus software 
 \cite{marques03,castro06,andrade15}.
We first compute the electronic ground state using spin-polarized density functional theory.
The spin-dependent Kohn-Sham orbitals $\psi_{si}(\mathbf{r})$ (indexed by spin $s=\uparrow,\downarrow$ and principal index $i$ in order of increasing energy starting at $i=1$) are eigenfunctions of the corresponding (effective) one-body Hamiltonians containing Kohn-Sham potentials,
\begin{align}
\label{eq:gsks}
    \hat{H}_s = -\frac{1}{2}\nabla^2 + V_s(\mathbf{r}).
\end{align}
The spin-dependent Kohn-Sham potentials
\begin{equation}
\label{eq:ksv}
    V_s(\mathbf{r}) = V_s(\mathbf{r};\rho_\uparrow,\rho_\downarrow) 
\end{equation}
are functionals of the electronic spin-density,
\begin{equation}
    \rho_s(\mathbf{r}) =\sum_{i=1}^{N_s} |\psi_{si}(\mathbf{r})|^2,
    \label{eq:ksrho}
\end{equation}
where $N_s$ is the number of electrons with spin $s$.
Since we use Troullier-Martins pseudopotentials to describe the core 1s electrons, we set $N_\uparrow=5$ and $N_\downarrow=4$ for a total of 9 valence electrons.
Furthermore we use  the Perdew-Zunger self-interaction correction implemented with Optimized Effective Potential model (OEP-KLI) \cite{OEP}.
The set of coupled equations (\ref{eq:gsks}-\ref{eq:ksrho}) is solved self-consistently for both the potentials $V_s(\mathbf{r})$ and the orbitals $\psi_{si}(\mathbf{r})$.

Next the time dependent calculations for N$_2^+$ are performed by solving a separate TDSE for each Kohn-Sham orbital,
\begin{equation}
    i\frac{\partial}{\partial t}\psi_{si\mathcal{E}}(\mathbf{r},t) = \left[\hat{H}_{s} + \mathcal{E}(\hat{x}\cdot\mathbf{r})f(t) \right]\psi_{si\mathcal{E}}(\mathbf{r},t), 
\end{equation}
with the initial wavefunction
\begin{equation}
    \psi_{si\mathcal{E}}(\mathbf{r},0) =\psi_{si}(\mathbf{r}),
\end{equation}
where the electric field has the same form as Eq. (\ref{eq:efield1D}) except with different parameters: the central frequency is tuned to the energy difference between the occupied $\psi_{\downarrow 2}$ orbital ($\sigma_u$ symmetry, energy $E_{\downarrow 2}=-1.0556$ a.u.) and the unoccupied $\psi_{\downarrow 5}$ orbital ($\sigma_g$  symmetry, energy $E_{\downarrow 5}=-0.9526$ a.u.), which in our model is at $\omega_0=0.1030$ (corresponding to 446 nm); we also use half the pulse duration of the 1D model, $\tau=200$ a.u., $t_0=2\tau=400$ a.u.
The same parameters as the 1D calculation are used for the grid and time-propagation, except that the box is now a cylinder of radius 20 a.u. and half-length 30 a.u. (still with a complex absorbing potential in the last 5 a.u. in each direction).
The time-dependent dipole moment is computed as a sum over all occupied orbitals,
\begin{align}
    D(\mathcal{E},t) = \sum_{s=\uparrow,\downarrow}\sum_{i=1}^{N_s} |\psi_{si\mathcal{E}}(\mathbf{r},t)|^2(\hat{x}\cdot\mathbf{r})d^3\mathbf{r},
\end{align}
(since both the molecular axis and the laser polarization are oriented along the $x$-axis, the dipole moment is only in that direction).
For the Fourier transform over time, we continue to use Eq. (\ref{eq:fttime}).

\begin{figure}
    \centering
    \includegraphics[width=0.6\linewidth]{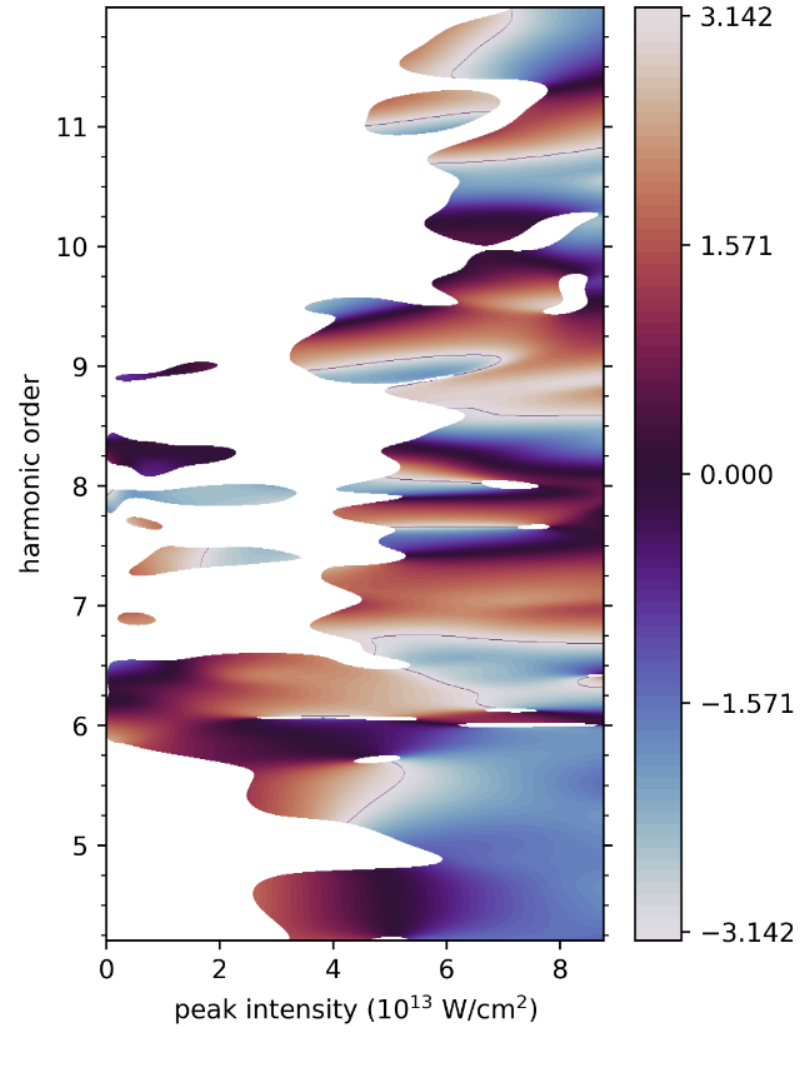}
    \caption{Microscopic calculations harmonics phases for N$_2^+$ for different peak laser intensities (the rest of the parameters as in Figure \ref{fig:N2micro}). 
}
    \label{fig:N2microphase}
\end{figure}

    \label{fig:N2microgdelay}

\begin{figure}
    \centering
    \includegraphics[width=0.6\linewidth]{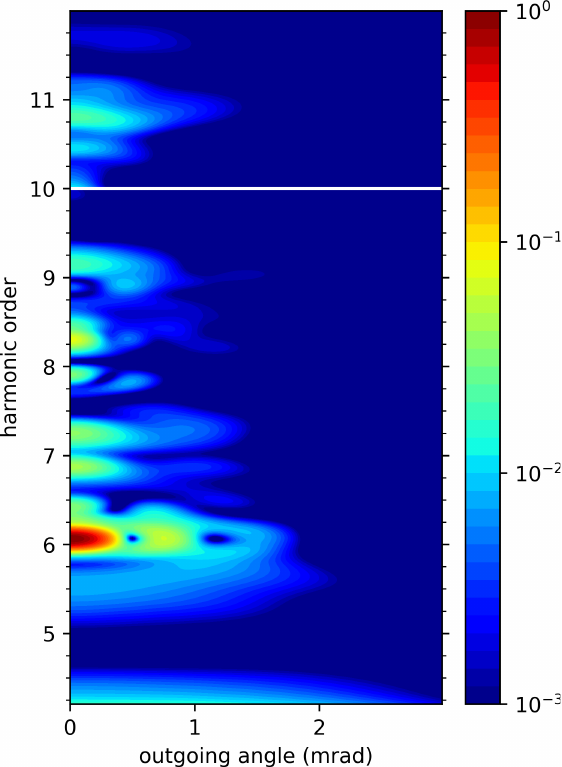}
    \caption{Results of macroscopic calculations: angle-resolved HHG spectrum for a macroscopic gas of aligned N$_+^2$  molecules.
    The horizontal axis is the outgoing angle  of the radiation relative to the laser axis, and the signal is
essentially zero for angles larger than 3 mrad. The spectrum at the highest frequencies (above the solid white line) has been scaled up by a factor of 10 so that it is visible. 
    }
    \label{fig:N2macro}
\end{figure}

\subsection{Interpolation and macroscopic response}
\label{sec:methods2}
The main difficulty in evaluating the integral in Eq. (\ref{eq:hdef}) is that the Bessel function is rapidly oscillating when $\kappa$ is moderately large, but it is prohibitively expensive to sample $D(\mathcal{E},\omega)$ tightly enough in $\mathcal{E}$ to resolve those oscillations.
One way to evaluate such an integral is to interpolate $D(\mathcal{E},\omega)$ from a small number of samples, then integrate the interpolant with enough points to resolve the oscillations of the Bessel function.
Leaving both the interpolation and integration steps completely general, the procedure is represented by the following equation,
\begin{eqnarray}
\label{eq:h_numeric}
    K(\kappa_i,\omega)& =& \lim_{n,m\rightarrow\infty}
     \sum_{j=0}^{n-1}\sum_{k=0}^{m-1}w_j^{(n)}J_0\left[\kappa_iq_j^{(n)}\right]
 \nonumber \\
&\times& I^{(m)}_{k}\left\{e^{-\left[q_j^{(n)}\right]^2}\right\} D\left[\mathcal{E}_0x^{(m)}_k,\omega\right]
\end{eqnarray}
where $\{\kappa_i\}$ are the values of $\kappa$ at which we would like to evaluate $K(\kappa,\omega)$ (note, we use the same $\kappa_i$ for all $\omega$),
$\left\{q^{(n)}_j,w^{(n)}_j\right\}$ is any family of integration nodes and weights such that
\begin{align}
    \int_0^\infty f(q)qdq = \lim_{n\rightarrow\infty} \sum_{j=0}^{n-1} w^{(n)}_jf(q^{(n)}_j),
\end{align}
for all smooth functions $f$ with compact support,
and $\left\{x^{(m)}_k,I^{(m)}_k(x)\right\}$ is any family of interpolation nodes and functions such that
\begin{align}
    f(x) = \lim_{m\rightarrow\infty} \sum_{k=0}^{m-1}I^{(m)}_{k}(x)f(x^{(m)}_k), \quad 0\leq x\leq 1
\end{align}
for all smooth functions $f$.
Throughout this section $j$ ranges from 0 to $n-1$ and $k$ ranges from 0 to $m-1$ even when we do not write these ranges explicitly.
Note that when evaluating Eq. (\ref{eq:h_numeric}) it is usually much faster to do the sum over $j$ before the sum over $k$, even though the opposite order is possibly more intuitive (interpolate then integrate).

For integration, the simplest option is to use the trapezoidal rule with uniformly spaced nodes,
$ q^{(n)}_j =  j/\sqrt{n},$ and 
    $w^{(n)}_j = j/n$.
The factors of $n^{-1/2}$ ensure that as $n\rightarrow\infty$ the range of integration extends to $[0,\infty]$ and at the same time the spacing goes to zero.
Unfortunately, the trapezoidal rule converges slowly so we use instead a more efficient integration method based on the roots of the Bessel function \cite{ogata2005},
    $q^{(n)}_j = r_j/\pi\sqrt{n}$,
    $w^{(n)}_j = 2/\pi^2nJ_1(r_j)^2$
where $r_j$ are the positive roots of $J_0$ in ascending order, and $J_1$ is the first-order Bessel function.
We have chosen the factors such that uniform nodes scheme reduces to Bessel roots scheme when $j\gg 1$ (because $r_j\rightarrow \pi j$ and $J_1(r_j)^2\rightarrow2/(\pi^2 j)$) meaning these two methods essentially only differ at the first few points.

\begin{figure}
    \centering
    \includegraphics[width=0.6\linewidth]{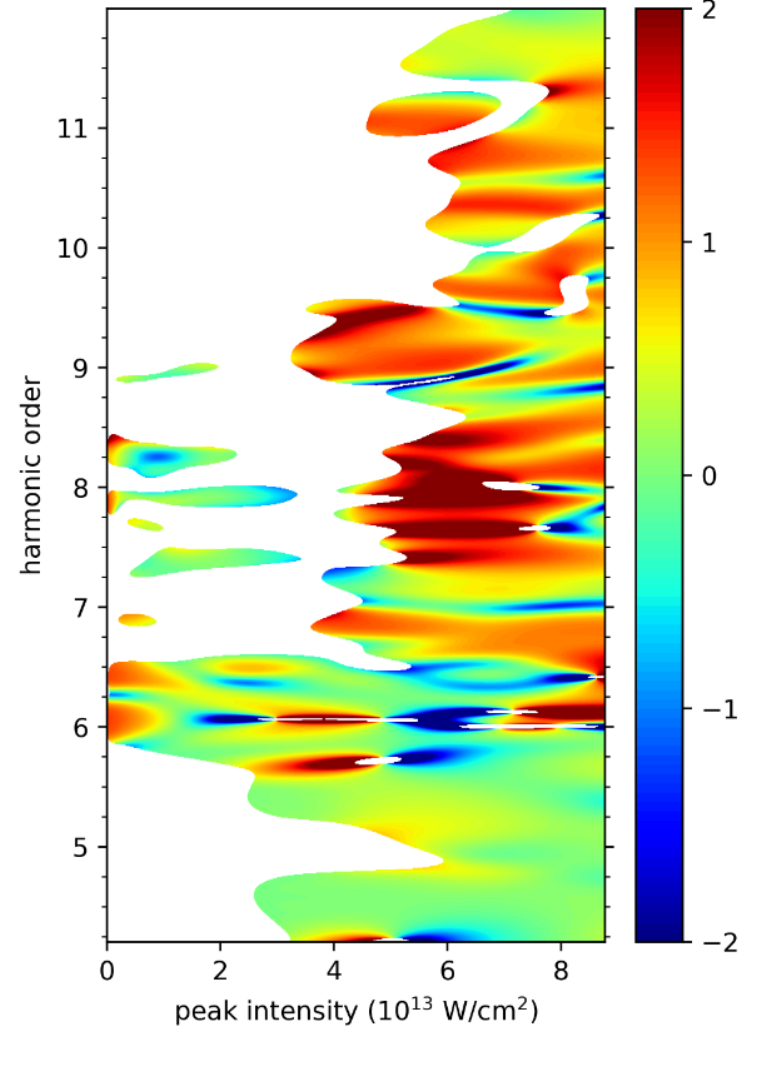}
    \caption{Microscopic calculations for harmonics group delay (in units of optical cycles) for N$_2^+$ 
    as a function of laser peak intensity.}
    \label{fig:N2microgdelay}
\end{figure}

For interpolation, we begin by defining the functions $\tilde{I}^{(L)}_{l}(x)$ which are polynomial interpolants for the Chebyshev nodes of the second kind on the interval $[-1,1]$.
In barycentric form,
\begin{align}
    \tilde{I}^{(L)}_{l}(x) =& \frac{W^{(L)}_l (x-\tilde{x}^{(L)}_l)^{-1}}{\sum_{l=0}^{L-1}{W^{(L)}_l (x-\tilde{x}^{(L)}_l)^{-1}}},
\end{align}
where
\begin{align}
    \tilde{x}^{(L)}_l =& \cos(\pi l/L),\quad l=0,\ldots,L, \\
    W^{(L)}_l =& 
    \begin{cases}
    0.5(-1)^l, & l=0,L,\\
    (-1)^l, &  l=1,\ldots,L-1.
    \end{cases}
\end{align}
For symmetry reasons $D(\mathcal{E},\omega)$ is an odd function of $\mathcal{E}$, which means we can perform an interpolation of order $L=2m$ using only $m$ sample points by choosing $x^{(m)}_k=\tilde{x}^{(2m)}_k$ for $k=0,\ldots,m-1$ and using symmetry to determine the remaining $m+1$ function values.
This consideration leads to the following expressions for the interpolation functions and nodes, which are what we use in Eq. \ref{eq:h_numeric},
\begin{align}
    I^{(m)}_k(x) =& \tilde{I}^{(2m)}_{k}(x) - \tilde{I}^{(2m)}_{2m-k}(x)\\
    x^{(m)}_k =& \cos(\frac{\pi k}{2m}).
\end{align}

Using these methods for integration and interpolation, we have studied the convergence of Eq. \ref{eq:h_numeric} with respect to $m$ and $n$ and found that $n=1000$ and $m=40$ are sufficient to converge the macroscopic HHG spectra of both the 1D and the 3D models and for this parameters the rest of the results have been computed.  
Fig. (\ref{fig:convergence}) shows the dependence result for macroscopic calculations for harmonics as a function of peak electric field  for interpolation based on 5, 10, 20 and 40 microscopic calculations respectively.

\begin{figure}
    \centering
    \includegraphics[width=0.6\linewidth]{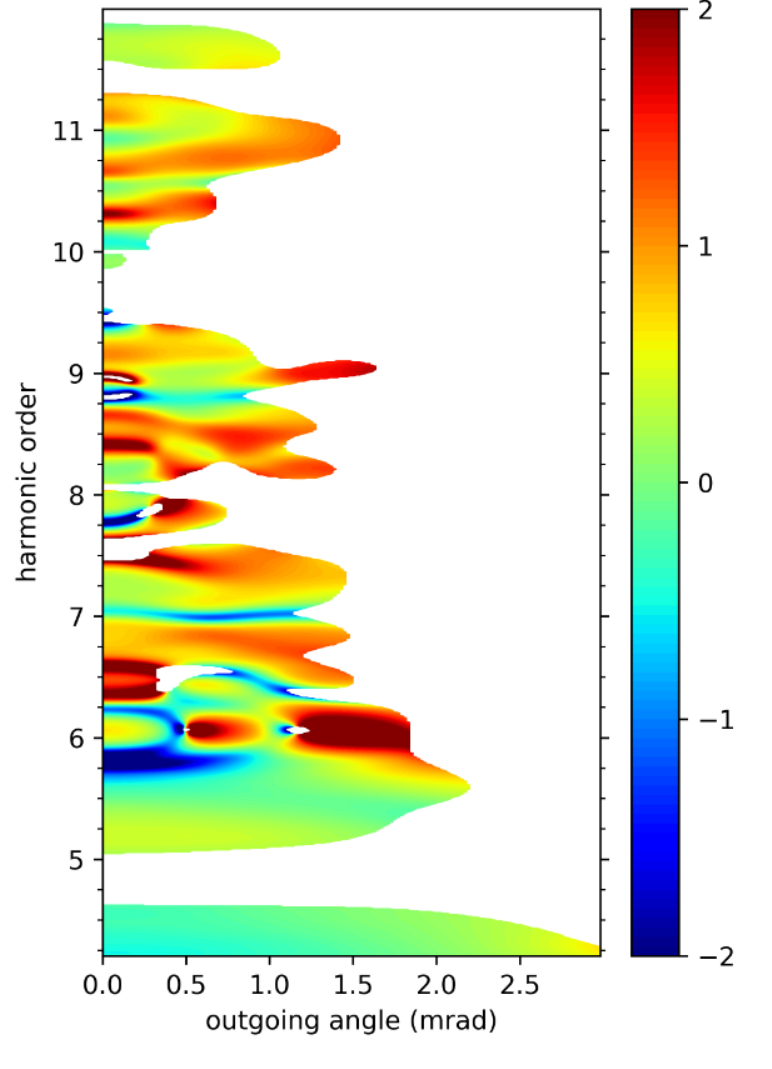}
    \caption{Results for macroscopic calculations harmonics group delay for N$_2^+$
(in units of optical cycle.)}
    \label{fig:N2macrogdelay}
\end{figure}
\begin{figure}
    \centering
    \includegraphics[width=0.6\linewidth]{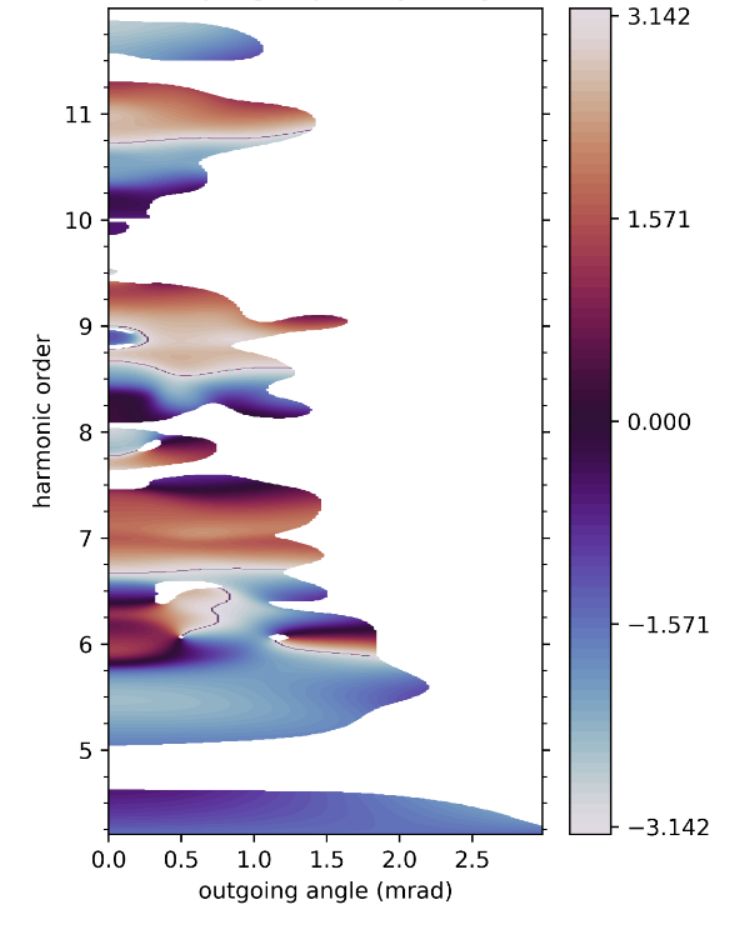}
    \caption{Results for macroscopic calculations harmonics  phases  for N$_2^+$
    }
    \label{fig:N2macrophase}
\end{figure}

\subsection{Floquet theory}
This section presents discussion of Floquet theory as it applies to HHG, because it can be used to understand the properties of the Mollow sidebands which appear in the single-molecule harmonic spectra.
In the long-pulse limit $\tau\rightarrow\infty$, the TDSE becomes
\begin{align}
\label{eq:floquet_tdse}
    i\frac{\partial}{\partial t}\ket{\Psi_\mathcal{E}(t)} = & \left[\hat{H} + \mathcal{E}\cdot\mathbf{r} \cos(\omega_0 t)\right]\ket{\Psi_\mathcal{E}(t)},
\end{align}
which is exactly $T$-periodic in time, where $T=2\pi/\omega_0$.
Although the general solution to Eq. (\ref{eq:floquet_tdse}) need not be $T$-periodic, the Floquet theorem does allow us to write it as a linear combination of $T$-periodic functions $\phi_{\mathcal{E}j}(t)$ with associated quasi-energies $E_{\mathcal{E}j}$,
\begin{align}
\label{eq:floquet_series}
    \ket{\Psi_\mathcal{E}(t)} = \sum_j c_j e^{-iE_{\mathcal{E}j}t}\ket{\phi_{\mathcal{E}j}(t)}.
\end{align}
The quasienergies are complex numbers defined modulo $\omega_0$ with the imaginary part related to the lifetime of the state $\tau_{\mathcal{E}j} = -1/\text{Im}[E_{\mathcal{E}j}]$ and the real part being the AC Stark-shifted energy of state $j$ in the presence of the laser field. As $\mathcal{E}\rightarrow 0$ the quasienergies approach the eigenvalues of $\hat{H}$ (modulo $\omega_0$) and the lifetimes of bound states approach $\infty$.
In this paper we do not consider the problem of determining the Floquet states and quasienergies numerically---we merely use Floquet theory as an analytical tool to interpret our results.

The coefficients $c_j$ in Eq. (\ref{eq:floquet_series}) depend on the initial state of the system before the electric field was switched on, and they also depend on how exactly the electric field was ramped up from $0$ to $\mathcal{E}$.
Most commonly, the system begins in the ground state $j=0$ and the field is switched on gradually, in which case the adiabatic theorem implies only a single Floquet state is occupied: $c_j = \delta_{j0}$ (where $\delta$ is the Kronecker delta function).
If this is true, the time-dependent dipole moment, 
\begin{align}
    \mathbf{d}(t) =&  \bra{\Psi_\mathcal{E}(t)}\mathbf{r}\ket{\Psi_\mathcal{E}(t)}= \bra{\phi_\mathcal{E}(t)}\mathbf{r}\ket{\phi_\mathcal{E}(t)},
\end{align}
will also be $T$-periodic, because $\ket{\phi_\mathcal{E}(t)}$ is $T$-periodic, which finally implies that the HHG spectrum (proportional to the Fourier transform of $d(t)$) only contains frequencies that are integer multiples of the fundamental frequency $\omega_0=2\pi/T$.

The adiabatic theorem can break down, however, near a so-called exceptional point or avoided crossing \cite{Rotter} when two Floquet states become close in energy,
\begin{align}
    E_{\mathcal{E}_cj}\approx E_{\mathcal{E}_ck} \quad (\text{modulo} ~\omega_0),
\end{align}
near some critical value of the field strength $\mathcal{E}_c\leq\mathcal{E}$ that is passed during the ramp-up.
In our case, we intentionally cause the adiabatic theorem to break down by choosing the laser frequency $\omega_0$ to be resonant with the transition between the initial ground state $j=0$ and the first excited state $j=1$.
This creates an exceptional point at $\mathcal{E}_c=0$ that is unavoidable as the laser ramps up.
As a consequence, two Floquet states $\ket{\phi_{\mathcal{E}0}(t)}$ and $\ket{\phi_{\mathcal{E}1}(t)}$ rather than one are required to properly describe the HHG spectrum, and wavefunction for the system can be written as superposition
\begin{equation}
    \ket{\Psi_\mathcal{E}(t)} = c_0 e^{-iE_{\mathcal{E}0}t}\ket{\phi_{\mathcal{E}0}(t)}+ c_1 e^{-iE_{\mathcal{E}1}t}\ket{\phi_{\mathcal{E}1}(t)},
\end{equation}
which means the dipole moment can be written as
\begin{eqnarray}
    \mathbf{d}(t) 
    &=& |c_0|^2\mathbf{d}_{00}(t) + |c_1|^2\mathbf{d}_{11}(t) \nonumber \\
    &+&\text{Re}\left[c_0^*c_1 e^{-i(
    E_{\mathcal{E}1}-E_{\mathcal{E}0})t}\mathbf{d}_{01}(t) \right],
\end{eqnarray}
where    
\begin{equation}    
    \mathbf{d}_{jk}(t) = \bra{\phi_{\mathcal{E}j}(t)}\mathbf{r}\ket{\phi_{\mathcal{E}k}(t)}.
\end{equation}
The first two terms are still $T$-periodic, so these contribute the usual odd harmonics.
The $\mathbf{d}_{01}(t)$ factor in the third term is also $T$-periodic, but it is multiplied by an additional time dependent factor $e^{\pm i(
E_{\mathcal{E}1}-E_{\mathcal{E}0})t}$.
The Fourier transform of the third term therefore produces sidebands at frequencies $n\omega_0 \pm (E_{\mathcal{E}1}-E_{\mathcal{E}0})$.
It is well known that the difference in quasienergies between two resonantly coupled states is equal to the Rabi frequency $E_{\mathcal{E}1}-E_{\mathcal{E}0} = \omega_{Rabi} = \mu\mathcal{E}$ where $\mu$ is the transition dipole moment.
The above analytical argument, as well as the numerical results of this paper and others, suggest that the sidebands should also appear in HHG spectra, although this has not been confirmed experimentally.

\section{Results and Analysis}
\label{sec:results}

Using the methods described Sec. \ref{sec:methods1} we have performed full TDSE calculations of single-molecule harmonic spectra for both a 1D model system and for N$_2^+$ in 3D as a function of the peak electric field strength up to $\mathcal{E}_0=0.03$ (peak intensity $I_0=3.16\times 10^{13}$ W/cm$^2$) in the first case and $\mathcal{E}_0=0.05$ (peak intensity $I_0=8.76\times 10^{13}$ W/cm$^2$) in the second case.
Next, the output of those TDSE calculations was used to compute the far-field harmonic spectra produced by macroscopic samples of such molecules, accounting for the fact that each molecule in the sample sees a different peak field strength based on its position within the Gaussian profile of the laser beam.
In the next subsection we present and analyze the results of those calculations.

\subsection{Single-molecule spectra}

Figure \ref{fig:1Dmicroa}  shows results of multiple calculations for different peak laser intensity (the rest of the laser parameters were kept the same) for 1D model of molecule. The analytic prediction $\omega=N\omega_0 \pm \mu$E for the sideband positions is indicated by the red dashed lines.
There Mollow sidebands are most pronounced around 5th and 7th harmonics for 1D molecule model.

Figure  \ref{fig:N2micro} presents TDDFT results for HHG spectra of a single aligned $N_2^+$ molecule driven by a 446 nm laser as a function of peak electric field amplitude while the lest of the laser pulse are the same. Besides the usual odd harmonics, many other features appear. All features that appear in addition to the odd harmonics are related either to Mollow sidebands or additional contributions from other excitations and interferences between different contributions.
As in the case of 1D model molecule, spectral features can be explained using an analytical model based on Floquet theory which are shown as dashed red lines.
The analytical model is very successful in predicting functional dependence for sidebands of low harmonics (1,3,5,6), and moderately successful for above-threshold harmonics (11,13). 
Features following the clear trend of intensity dependence of Mollow sidebands are also visible near the 7th and 9th harmonics for TDDFT calculations, but they are partially obscured by numerous irregular features, which we attribute to
Rydberg states (since these harmonics are just below the ionization threshold at 9.39$\omega_0$). These
irregular features are also signatures of nonadiabatic dynamics, albeit of a more complex type which goes beyond the simple two-level dressed state model. 

Finally, the strong feature around 6$\omega_0$ is caused by beating between two excited states and matches the energy difference between the   $2\sigma_{g,d}$ and $2\sigma_{u,d}$ molecular orbitals (shown as dotted blue line), and therefore represents the hole being driven further down in energy. 
This indicates more complicated 3-level laser-induced dynamics.
As the laser intensity increases, the 6$\omega_0$ 
feature bifurcates because the energies of the dressed states are shifting up and down symmetrically. (Note that the feature has a very strong macroscopic response.)
The 'splitting' of the 7$\omega_0$ peak may be caused by similar effect, given that the separation is around the same magnitude.

Figure \ref{fig:N2microphase} presents phases of the harmonics as a function of the peak laser intensity for TDDFT single molecule calculations. 
The phase of the harmonics is relatively constant, except for the sidebands on the 5th harmonic, which reverse sign once as a function of intensity. This can be accounted for by the phase of the Rabi oscillation at the peak of the laser pulse. For longer pulse duration, we would expect the phase to change more rapidly. As a function of frequency, the phase tends to increase. This indicates a time delay meaning many of the higher harmonics are produced slightly after the
peak, which is confirmed in Figure \ref{fig:N2microgdelay} presenting group delays for the harmonic spectrum. There are several nodes where the spectrum goes to zero, and the phase rotates through
 $2\pi$ around a single point. Most notably, several nodes are clustered together around the 6$\omega_0$ feature, reﬂecting a complicated intensity dependence.

Next we present the results of the macroscopic HHG calculations as described in preceding sections, obtained using the microscopic calculations.

\subsection{Macroscopic spectra}

Figure \ref{fig:1Dmicrob} shows results of multiple calculations for different peak laser intensity (the rest of the laser parameters were kept the same) for 1D model of molecule. 
The analytic prediction $\omega=N\omega_0 \pm \mu$E for the sideband positions is indicated by the red dashed lines.
There Mollow sidebands are most pronounced around 5th and 7th harmonics for 1D molecule model.

Figure \ref{fig:N2macro} shows the macroscopic HHG spectrum for a laser with a peak ﬁeld peak intensity 8.78×$10^{13}$ W/cm$^2$ at ﬁxed outgoing azimuthal angle $\phi=\pi/2$. The vertical axis is the outgoing angle $\theta$ (in milliradians). The color is again on a logarithmic scale truncated at 0.1 $\%$ the maximum
value of the spectrum. 

In the macroscopic spectrum  nearly every nonadiabatic feature
from the single-molecule spectrum is still present, including the sidebands around 5$\omega_0$ and 11$\omega_0$, the splitting of the 7$\omega_0$ peak, the more complicated Rydberg structure, and the strong feature around 6$\omega_0$. The bifurcation of the 6$\omega_0$ feature   in the macroscopic spectrum, leads to a complex angular pattern including several angular nodes. The angular pattern
of the Rydberg structure is also very intricate. 
At this point one might speculate
 that the angular dependence could be used to distinguish different types of features in the spectrum. 
For example sidebands radiate at larger angles than the main harmonics, and they bend slightly inward. This is a clear signature to look for in the macroscopic spectrum.

Figure \ref{fig:N2macrophase} presents results for macroscopic calculations for the harmonics phases as function of the outgoing angle. The dependence of the phases on the angle is rather weak except the feature around 6th harmonic. The spectrum that 
is obtained in the macroscopic calculations is related to the case when phases of the sidebands and the main harmonic are similar. Compared to microscopic result for the phases of the harmonics the macroscopic result is simpler and several features are 'washed out'.

Figure \ref{fig:N2macrogdelay} shows results for macroscopic calculations for the group delays of the harmonics (in units of optical cycle) as a function of the outgoing 
angle. The results show that the corresponding sidebands exhibit different group delay than the main harmonic. 

\section{Conclusions}

In conclusion, we have used a combination of ab initio quantum calculations and macroscopic wave-propagation solution to simulate high harmonic spectroscopy of N$_+^2$. Our results  indicate that signatures of complex nonadiabatic dynamics remain visible in the macroscopic spectra, and therefore should be observable under realistic experimental conditions. Furthermore, the angular patterns of these features in the far-ﬁeld are often highly nontrivial, suggesting they contain additional information about the nonadiabatic dynamics. Because nonadiabatic dynamics is a general feature of strong ﬁeld processes in open shell molecules. In principle similar results can be anticipated for many examples of molecular superposition states and our conclusions have potentially wider applicability. This work represents an important step towards understanding and controlling the nonadiabatic dynamics of molecules in strong laser ﬁelds. Additionally, the novel approach we have developed to interface macroscopic description of generated light,  with TDDFT (or other ab initio techniques) opens the door for macroscopic analysis of many other ultrafast phenomena and studies of more complex systems.  

\section{Funding}
A. J. acknowledge support from National Science Foundation PHY-2317149 and PHY-2110628 awards. 
Funding during the
final months of the project was provided by the Chemical Sciences,
Geosciences, and Biosciences Division, Office of Basic Energy Sciences,
Office of Science, U.S. Department of Energy, Grant no. DE-FG02-86ER13491.

Simulations were performed on the Alpine high-performance computing resource jointly funded by the University of Colorado Boulder, the University of Colorado Anschutz, and Colorado State University.

\end{document}